\documentclass[aip,jap,reprint,showpacs,twocolumn,superscriptaddress]{revtex4-1}
\usepackage{amsmath}
\usepackage{amssymb}
\usepackage{graphicx}
\usepackage{hyperref}
\usepackage{url}
\begin{document}
\title{Strain-engineered A-type antiferromagnetic order in YTiO$_3$: a first-principles calculation}
\author{Xin Huang}
\author{Yankun Tang}
\affiliation{Department of Physics, Southeast University, Nanjing 211189, China}
\author{Shuai Dong}
\affiliation{Department of Physics, Southeast University, Nanjing 211189, China}
\affiliation{Laboratory of Solid State Microstructures, Nanjing University, Nanjing 210093, China}
\date{\today}

\begin{abstract}
The epitaxial strain effects on the magnetic ground state of YTiO$_3$ films grown on LaAlO$_3$ substrates have been studied using the first-principles density-functional theory. With the in-plane compressive strain induced by LaAlO$_3$ (001) substrate, A-type antiferromagnetic order emerges against the original ferromagnetic order. This phase transition from ferromagnet to A-type antiferromagnet in YTiO$_3$ film is robust since the energy gain is about 7.64 meV per formula unit despite the Hubbard interaction and modest lattice changes, even though the A-type antiferromagnetic order does not exist in any $R$TiO$_3$ bulks.
\end{abstract}
\maketitle

%%-------------------------------------------------------------------------
%
%\section{INTRODUCTION}
Transition metal oxides with the perovskite structure exhibit a wide variety of electronic phases with plenty charge-, magnetic-, and orbital-structures, and show many prominent functionalities including colossal magnetoresistance, high-$T_C$ superconductivity,
and metal-insulator transitions.\cite{Dagotto:Sci,Tokura:Rpp} Among these oxides, the $R$TiO$_3$ family ($R$ is a rare earth cation), whose 3$d$ $t_{\rm 2g}$ bands lay near the Fermi level, have not been extensively studied. % show quite different physical properties from well-studied manganites or cuprates. In the past decades, manganites, cuprates, and ferrates have been extensively studied both experimentally and theoretically, while much less few attention has been paid to $R$TiO$_3$.
However, $R$TiO$_3$ is not a feature-less family, which also owns rich spin/orbital ordered phases.\cite{Mochizuki:Njp} These phases also involve the couplings between charge, orbital, lattice, and spin degrees of freedom, which have the potential to be used in spintronic or correlated-electron devices.%For example, very recently, the EuTiO$_3$ is predicted and further confirmed to be a multiferroic materials under certain strain conditions [lee:nature476/466][Rabe:PRL].

$R$TiO$_3$'s with trivalent $R$ cations are all protype Mott-Hubbard insulators and their common crystal structure is a pseudocubic perovskite with an orthorhombic distortion (the GdFeO$_3$-type distortion). This distortion arises from the tilting TiO$_6$ octahedron around the [$110$] axis and a followed rotation around the [$001$] axis. The magnitude of this distortion depends on the ionic radii of $R$.
Similar to the $R$MnO$_3$ case, the lattice structure is more distorted with a small $R$ and the Ti-O-Ti bond angle is decreased more significantly from $180^{\rm o}$. The GdFeO$_3$-type distortion plays a crucial role in controlling the subtle competitive exchange interactions in these insulating titanates. %, which is also analogous with $R$MnO$_3$.
The magnetic ground state of $R$TiO$_3$ exhibits a transition from ferromagnetic (FM) order to antiferromagnetic (AFM) one with increasing size of $R$ cation.\cite{Katsufuji:Prb,Greedan:Jlcm}
%According to Ref. [Mochizuki:jpsj:2], the FM exchange along the c-axis in compounds with small $R$ such as GdTiO$_3$ and YTiO$_3$ is realized by the super-exchange processes mediated by the $e_{\rm g}$ orbitals due to $t_{\rm 2g}$-$e_{\rm g}$ hybridizations induced by the GdFeO$_3$-type distortion[Mochizuki:jpsj:2] and the FM coupling in the $ab$-plane can be understood straightforwardly as the ferromagnetism with antiferro-orbital ordering. These studies emphasized the effect of orbital ordering to magnetic coupling in $R$TiO$_3$ which was experimentally confirmed in these compounds. The orbital ordering is robust due to both the Jahn-teller distortion and $t_{\rm 2g}$-$e_{\rm g}$ hybridizations.

It is very interesting to compare the phase diagrams of $R$TiO$_3$ and $R$MnO$_3$, %, as shown in Fig.~\ref{sketch}(b),
both of which show magnetic transitions with increasing size of $R$ cations. And the Curie (or N\'eel) temperatures show V-shape behaviors near the critical points in both families. However, there are also two key differences between $R$TiO$_3$ and $R$MnO$_3$. First, the FM-AFM tendency is opposite in these two families. With a smaller $R$, $R$MnO$_3$ is more AFM but $R$TiO$_3$ is more FM. Second, the phases revealed in $R$MnO$_3$ are more complex than those in $R$TiO$_3$. The AFM phase in $R$TiO$_3$ bulks is the simple G-type AFM one while the AFM phases (e.g. A-type AFM, spiral-spin order, E-type AFM) in $R$MnO$_3$ are more complex which can be more interesting than the simple G-type one.\cite{Dong:Prb08} Thus it is nontrivial to ask whether is there any more (hidden) magnetic orders in $R$TiO$_3$? In a previous theoretical work Ref.~\onlinecite{Mochizuki:Jpsj01}, total energies of different magnetic structures including A-type AFM, FM and G-type AFM were calculated by using an effective spin-pseudospin Hamiltonian, which showed that the A-type AFM to FM phase transition occurs with increasing GdFeO$_3$-type distortion while the G-type AFM one has much higher energies. However, this result disagrees with the experimental phase diagram since the G-type AFM phase is very robust in $R$TiO$_3$ with large $R$ while the A-type AFM phase has not been observed in any real $R$TiO$_3$ compounds so far.

In this paper, by using the first-principles calculations, we intend to investigate the effects of strain on magnetic structures of YTiO$_3$ film, focusing on the phase transition of the magnetic ground state. Our calculation predicts that a robust A-type AFM phase can be stabilized by an in-plane compressive strain by using small lattice substrates like LaAlO$_3$.

%The remaining parts of this paper are organized as follow. The procedure of the first-principles calculations is given in Sec. II. The main results and discussion on magnetic structure affected by strain are presented in Sec. III. This work is concluded in Sec. IV.

%%------------------------------------------------------------------------------
%
%\section{METHOD AND PROCEDURE OF FIRST-PRINCIPLES CALCULATIONS}

%A sketch of the lattice structure of YTiO$_3$ is presented in Fig.~\ref{sketch}(a).
YTiO$_3$ bulk has a orthorhombic structure (space group \textit{Pbnm}) with lattice constants of $a$=$5.358$ \AA{}, $b$=$5.696$ \AA{}, and $c$=$7.637$ \AA{}. Such a minimum unit cell consists of 4 formula units. To simulate the effect of in-plane compressive strain induced by the substrate, the lattice constants along the a-axis and b-axis are fixed to $a$=$b$=$5.366$ ($3.794\times\sqrt{2})$ \AA{} to match the
($001$) LaAlO$_3$ substrate. Here LaAlO$_3$ is adopted as the substrate to give a weak in-plane compressive strain to YTiO$_3$ film since the in-plane
lattice constant of LaAlO$_3$ is a little smaller ($\sim3\%$) than that of YTiO$_3$ itself. Such a small difference between lattice constants also promises
probable epitaxial growth of YTiO$_3$ thin films on LaAlO$_3$ substrate.

%\begin{figure}
%\centering
%\includegraphics[width=0.4\textwidth]{Graph1}
%\caption{(Color online) (a) Structure schema of YTiO$_3$. The red (small), blue (medium) and green (big) balls are the O, Ti and Y ions, respectively. Translucent area represents the TiO$_6$ octahedra. (b) The magnetic phase diagram for perovskite $R$TiO$_3$ family (Upper panel) and $R$MnO$_3$ family (Lower panel).}
%\label{sketch}
%\end{figure}

Our first-principles calculations were performed using density-functional theory (DFT) within the generalized gradient approximation GGA+U method\cite{Blochl:Prb2,Kresse:Prb99} with the Perdew-Becke-Erzenhof parametrization\cite{Perdew:Prl} as implemented in the Vienna \emph{ab} initio simulation package (VASP).\cite{Kresse:Prb,Kresse:Prb96} The valence states include $4d^15s^2$, $3d^24p^2$ and $2s^22p^4$ for Y, Ti and O, respectively. The lattice optimization and all other static computations have been done with the Hubbard $U$ on the $d$-electrons of Ti$^{3+}$ ion, and the Dudarev\cite{Dudarev:Prb} implementation with $U_{\rm eff}$ = $3.2$ eV has been used if not noted explicitly.\cite{Sawada:Prb}. The atomic positions are fully optimized as the Hellman-Feynman forces are converged to less than $1.0$ meV/\AA{}. This optimization and the electronic self-consist iterations are performed using the plane-wave cutoff of $500$ eV and a $9\times9\times6$ Monkhorst-Pack $k$-point mesh\cite{Monkhorst:Prb} centered at $\Gamma$ grid in combination with the tetrahedron method.\cite{Blochl:Prb}

%%------------------------------------------------------------------------
%

%\section{RESULTS AND DISCUSSTION}

%\subsection{Magnetic ground state}

First, the ground state of YTiO$_3$ bulk is checked. Using the experimental crystal structure, non-magnetic (NM) state and four magnetic orders: FM, A-type AFM, C-type AFM and G-type AFM, have been calculated to compare the energies.
Within GGA+$U$, our calculations confirm that the FM order has the lowest energy and the calculated local magnetic moment is $0.88\mu_B$/per Ti in agreement with the experimental magnetic moment ($0.84\mu_B$).\cite{Garrett:Mrb}
The detail results of calculated total energy are summarized in Table I. According to Table~\ref{table}, %the ground state is clearly a robust FM state, since
other magnetic orders' energies (per Ti) are higher than the FM one: $3$ meV higher for A-type AFM, $6$ meV higher for C-type AFM and $7$ meV higher for G-type AFM. %The A-type AFM is the closest one to FM state and the difference between FM and G-type AFM is the biggest.
It should be noted that the FM ground state is robust within a large region of $U_{\rm eff}$ from $0$ eV to $5$ eV (not shown here). Thus, our calculations agree quite well with the experimental results and previous DFT studies.\cite{Sawada:Prb,Sawada:Pb}

\begin{table}
\caption{The energy difference $\Delta_E$ (per Ti) between magnetic states and the NM state for unstrained bulk YTiO$_3$: $E$(magnetic)-$E$(NM), in unit of eV and the corresponding local magnetic moments per Ti in unit of $\mu_B$.}
\begin{tabular*}{0.48\textwidth}{@{\extracolsep{\fill}}llllllr}
\hline \hline
Magnetic order & NM & FM & A-AFM & C-AFM & G-AFM \\
\hline
$\Delta E$ & $0$ & $-0.533$ & $-0.530$ & $-0.527$ & $-0.526$\\
Magnetic moment & $0$ & $0.88$ & $0.86$ & $0.83$ & $0.82$\\
\hline \hline
\end{tabular*}
\label{table}
\end{table}

\begin{figure}
\centering
\includegraphics[width=0.5\textwidth]{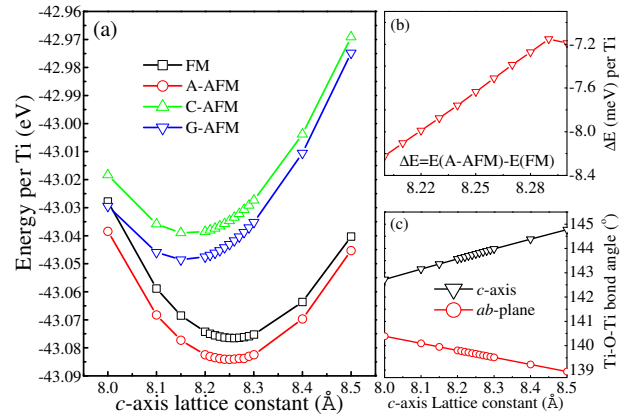}
\vskip -0.5cm
\caption{(Color online) (a) Energies for different magnetic orders as a function of the $c$-axis lattice constant. %The blue, green, black and red lines represent C-type AFM, G-type AFM, FM and A-type AFM, respectively.
(b) The energy difference between the A-type AFM and FM as a function of the $c$-axis lattice constant. (c) The Ti-O-Ti bond angle in $ab$-plane and along the $c$-axis respectively for the A-type AFM state.} %All these calculations are performed in YTiO$_3$ films with epitaxial strain.}
\label{energy}
\end{figure}

Subsequently, DFT calculations with the epitaxial strain are performed. Epitaxial strain is here realized by fixing the in-plane lattice constants to fit the LaAlO$_3$ substrate as stated before, while the lattice constant along c-axis is varied from $7.0$ \AA{} to $9.0$ \AA{} to search the equilibrium one under the strain, as shown in Fig.~\ref{energy}(a). In our calculations, the internal atomic positions are relaxed with magnetism under each fixed lattice framework to obtain optimal crystal structures for calculating accurate energies. According to Fig.~\ref{energy}(a), it is obvious that the C-type AFM and G-type AFM states are much higher in energy than the FM and A-type AFM states. Thus, in the following, we will mainly focus on the FM and A-type AFM states. % which's energies are lower.
The relaxed lattice constant along $c$-axis is $8.25$ \AA{} for the A-type AFM state and $8.26$ \AA{} for the FM state. These two values are very close, implying that the magnetostriction is weak in YTiO$_3$, at least along the c-axis. And with the optimized $c$-axis lattice constant, the A-type AFM state has a lower energy than FM one, e.g. the energy difference between FM and A-type AFM reaches $7.64$ meV per Ti. %Comparing with the well known FM ground state in bulk YTiO$_3$,
The A-type AFM state appeared in strained YTiO$_3$ films is quite nontrivial since it does not exist in any $R$TiO$_3$ bulk, namely it is a new phase for $R$TiO$_3$ family. More importantly, this strain-induced phase transition from FM to A-type AFM is quite promising according to our calculation. As shown in Fig.~\ref{energy}(b), the energy difference between these two orders does not change sign for a large range of $c$-axis lattice constant around the optimized one, which means this transition is not sensitive to the optimized $c$-axis lattice constant. Noted that the energy difference is relatively significant since the energy difference in bulk YTiO$_3$ is only $3$ meV per Ti. To confirm that this phase transition is robust against the change of Hubbard parameter, the energy difference between FM and A-type AFM states is calculated with different $U_{\rm eff}$ from $0$ eV to $5$ eV stepped by $1$ eV, which changes from $16$ meV to $2$ meV (always positive). In other words, this FM to A-type AFM transition will not change by varying the Hubbard interaction $U_{\rm eff}$ in a large value (from $0$ eV to $5$ eV) In short, this strain-induced A-type AFM phase will be very promising to be found in real thin films even if the experimental lattice constant along $c$-axis and its Hubbard interaction are not exactly the same with those in our calculations.

To understand the underlying physical mechanism, it is meaningful to compare the Ti-O-Ti bond angle in YTiO$_3$ with and without the strain, as shown in Table.~\ref{bondangle}. %And the overall relation between the Ti-O-Ti bond angle for the strained YTiO$_3$ and the variable $c$-axis lattice constant is showed in Fig.~\ref{energy}(c).
According to Fig.~\ref{energy}(c), the bond angle in the $ab$-plane decreases but the one along $c$-axis increases with the increasing $c$-axis. These results imply that YTiO$_3$ is compressed and thus more distorted in the $ab$-plane but elongates and thus is less distorted along the $c$-axis.

\begin{table}
\caption{Bond angles in the $ab$-plane and along $c$-axis of YTiO$_3$ film on LaAlO$_3$ substrate and bulk YTiO$_3$.} %Here, both A-type AFM and FM orders in YTiO$_3$ film are compared at their equilibrium positions, respectively. The bulk YTiO$_3$ show ferromagnetic order.}
\begin{tabular*}{0.48\textwidth}{@{\extracolsep{\fill}}llllr}
\hline \hline
Ti-O-Ti & YTiO$_3$ film & YTiO$_3$ film & bulk YTiO$_3$ \\
bond angle & (A-AFM) & (FM) & (FM)\\
\hline \hline
$ab$-plane & $139.6^{\rm o}$ & $139.5^{\rm o}$ & $144.3^{\rm o}$\\
$c$-axis & $143.7^{\rm o}$ & $143.5^{\rm o}$ & $141.9^{\rm o}$\\
\hline \hline
\end{tabular*}
\label{bondangle}
\end{table}

%In strained films, the Ti-O-Ti bond angle in $ab$-plane decreases comparing with bulk YTiO$_3$, while the bond along $c$-axis becomes more straight. This bond-angle change is helpful to understand the magnetic transition.
As stated before, it is well known that in $R$TiO$_3$ compounds, small Ti-O-Ti bond angles with more distorted lattice structure exhibit FM order while those with larger Ti-O-Ti bond angle tend to be AFM. It should be noted that this tendency is in opposite to the tendency in manganites. In manganites, the straighter (closer to $180^o$) bond angle should lead to ferromagnetic exchange. Therefore, the strain driven phase transition here is not exactly the same with the corresponding one in strained manganites.\cite{Lee:Prl10} So here in strained case,the spin order in $ab$-plane has a tendency towards FM correlation due to decreasing in-plane Ti-O-Ti bond angle, while the increasing bond angle along $c$-axis tends to drive spins arranged antiparallel. In this sense, the emergence of A-type AFM in YTiO$_3$ on LaAlO$_3$ substrate can be qualitatively understood as the ferromagnetism with decreasing bond angle in $ab$-plane and antiferromagnetism with increasing bond angle along $c$-axis.
Of course, a comprehensive understanding of this magnetic transition needs more careful studies from microscopic theory.
Furthermore, the bond angles are very close between FM and A-AFM states in strained films, implying that the exchange-striction effect is weak in this materials, which is different from the strong exchange-striction in manganites\cite{Picozzi:Prl} or ion-selenides.\cite{Lw:Prb}
%In short, the lattice distortions in YTiO$_3$ film are mainly own to the strain induced by LaAlO$_3$ substrate, and then the magnetic ground state is influenced by the lattice distortion.

%%------------------------------------------------------------------------
%

%\subsection{Density of states and band structures}

\begin{figure}
\centering
\includegraphics[width=0.5\textwidth]{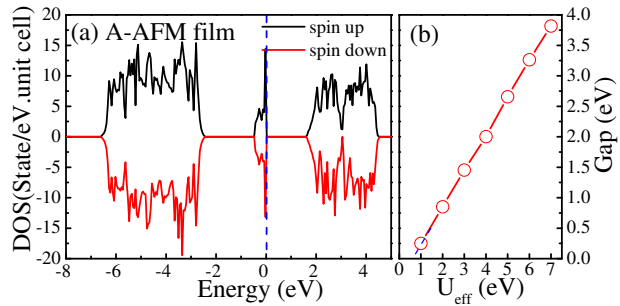}
\vskip -0.5cm
\caption{(Color online) (a)Total DOS of YTiO$_3$ film. The Fermi energy is positioned at zero. (b) The energy gap as a function of $U_{\rm eff}$. The critical $U_{\rm eff}$ value for zero gap is estimated as $0.6$ eV by extrapolation.}
\label{DOS}
\end{figure}

In many correlated electron materials, accompanying magnetic transitions, conductance often changes drastically, e.g. metal-insulator or insulator-superconductor transitions. Therefore, it is necessary to check the density of states (DOS) of YTiO$_3$ under strain, as shown in Fig.~\ref{DOS}(a).

First of all, YTiO$_3$ bulk is an insulator in our DFT calculation (not shown here), in agreement with the experimental result.\cite{Mochizuki:Njp} This insulating behavior is due to the Coulomb repulsion between $3d$ electrons, implying a Mott insulator. This scenario can be easily demonstrated because the pure GGA calculation without the Hubbard $U$ gives a metallic DOS. The band gap is $1.504$ eV for bulk YTiO$_3$, a little overestimated compared with the experimental data $1.2$ eV.\cite{Okimoto:Prb} Our calculation found a gap as $1.564$ eV for YTiO$_3$ under strain, which is only a little larger than the bulk value. This gap exists within a wide range of $U_{\rm eff}$ from $1$ eV to $7$ eV, though the detail value depends on $U_{\rm eff}$, as shown in Fig.~\ref{DOS}(b). Therefore, it is safe to say that the strained YTiO$_3$ remains an insulator and its conductance is not obviously changed since the gaps are almost identical.

%%------------------------------------------------------------------------
%

%\section{SUMMARY}

In summary, we have studied the effects of epitaxial strain on the magnetic ground states in YTiO$_3$ films. Our results predicted a new magnetic ground phase A-type antiferromagnet which had not been realized in any $R$TiO$_3$ bulk compounds. This robust A-type AFM phase is stabilized by an in-plane compressive strain induced by LaAlO$_3$ substrate. Its origin is understood as the ferromagnetism with decreasing bond angle in the \emph{ab}-plane and antiferromagnetism with increasing bond angle along the $c$-axis. Furthermore, the density of states calculation confirmed that the insulating behavior and the energy gap would not be significantly affected by this strain driven magnetic transition.

%\begin{acknowledgments}
We thank X. Z. Lu, H. J. Xiang, Q. F. Zhang, H. M. Liu for helpful discussion. Work was supported by the 973 Projects of China (Grant No. 2011CB922101) and NSFC (Grant Nos. 11004027 and 11274060).
%\end{acknowledgments}


\end{document}